\documentclass[twocolumn,10pt]{article}

\pdfoutput=1

\usepackage[latin1]{inputenc}
\usepackage{amsmath}
\usepackage{graphicx}
\usepackage{float}
\usepackage{multirow}

\DeclareMathOperator{\erf}{erf}

\makeatletter
\let\@fnsymbol\@arabic
\makeatother

\title{Error Estimation, Error Correction and Verification In Quantum Key Distribution}
\author{Øystein Marøy\thanks{Department of Electronics and Telecommunications, Norwegian University of Science and Technology, Trondheim, Norway. Electronic address: \it{oystein.maroy@iet.ntnu.no}}, Magne Gudmundsen\thanks{Department of Physics, Norwegian University of Science and Technology, Trondheim, Norway}, Lars Lydersen\footnotemark[1], Johannes Skaar\footnotemark[1]}

\begin{document}

\maketitle

\begin{abstract}

We consider error correction in quantum key distribution. To avoid that Alice and Bob unwittingly end up with different keys precautions must be taken. Before running the error correction protocol, Bob and Alice normally sacrifice some bits to estimate the error rate. To reduce the probability that they end up with different keys to an acceptable level, we show that a large number of bits must be sacrificed. Instead, if Alice and Bob can make a good guess about the error rate before the error correction, they can verify that their keys are similar after the error correction protocol. This verification can be done by utilizing properties of Low Density Parity Check codes used in the error correction. We compare the methods and show that by verification it is often possible to sacrifice less bits without compromising security. The improvement is heavily dependent on the error rate and the block length, but for a key produced by the IdQuantique system Clavis$^2$, the increase in the key rate is approximately 5 
percent. We also show that for systems with large fluctuations in the error rate a combination of the two methods is optimal. 
\end{abstract}

\section*{Introduction}
Quantum Key Distribution (QKD) \cite{bennett1984} is a method to distribute a secret key between two parties, Alice and Bob, through a quantum channel. An eavesdropper Eve is allowed full control over the channel. After the communication through the quantum channel Alice and Bob reconcile their keys using an error correction protocol. Using a privacy amplification protocol \cite{bennett1988, bennett1995} any information Eve might have about the key is removed. The unconditional security of the entire protocol can be proven using the laws of quantum mechanics \cite{shor2000, koashi2005, koashi2009}.

For practical QKD, the secret key rate is an important factor. The main limitations on the key rate is the transmission efficiency of the quantum channel and the performance of detectors at the receiving end of the channel, especially detector dead time. Developing better equipment is therefore important for making QKD a viable alternative for secure communication. However it is also possible to increase the key rate by more efficient error correction and privacy amplification protocols.

Due to imperfect equipment and Eves possible actions during the distribution phase, errors between Alice and Bobs keys are inevitable. Thus they need to do error correction, ending up with identical keys. This is done by classical communication on an authenticated channel. Because this communication reveals some information about the key, either the communication must be encrypted using previously established key, or additional privacy amplification must be used. Thus it is important to have an effective error correction protocol, revealing as little information about them as possible. Assuming a block of $N$ bits, containing $N\delta$ errors, the number of bits $L$ lost in error correction is lower bounded by the Shannon limit \cite{shannon1948}. For a perfect protocol, working at the Shannon limit we have
\begin{equation} \label{Shannon} 
L = N h(\delta)
\end{equation}
Here $h(\cdot)$ is the binary entropy function $h(p)=-p\log p-(1-p)\log(1-p)$.

\section*{Error correction}
Error correction in QKD is generally done by exchange of parity information about Alice's and/or Bob's keys. For processing purposes the key is divided into blocks of $N$ bits, on which error correction is performed while the next block is distributed on the quantum channel. Different protocols can be used for error correction, the most popular being CASCADE \cite{brassard1994}.

Of significant interest are also protocols using Low Density Parity Check (LDPC) codes \cite{gallager1963, mackay1996}. Using the technique of Density Evolution \cite{richardson2001} it is possible to construct error correcting codes performing extremely close to the Shannon limit \cite{chung2001}. In addition to being efficient, error correction protocols based on LDPC has another advantageous property. Let $d_{\text{min}}$ be the minimal Hamming distance between two codewords in the code, i.e. the minimal number of bits flips needed to turn a codeword into another. Then Alice and Bob's keys differ in at least $d_{\text{min}}$ bits if the error correction protocol completes without beeing successful. Finding $d_{\text{min}}$ for a code is not solvable in polynomial time, but one can find a lower bound. A linear code cannot correct more errors than $\frac{d_{\text{min}}}{2}$. If the code performs at the Shannon limit this gives
\begin{equation}\label{dmin}
d_{\text{min}}=2N\delta.
\end{equation}
Note that for optimal efficiency a different code is needed for each error rate. Because creating good codes is computationally demanding, and therefore a time consuming task, a running QKD system would need an large set of preestablished codes, each optimized for a different error rate. 

Both CASCADE and LDPC based protocols require an estimate on the error rate. This error estimation is often done by random sampling. Alice and Bob publicly announce some random bit pairs from their keys to estimate the error rate. However, the estimation can also be done without sacrificing bits. For example, in both protocols the error rate of the previous block is known to Alice and Bob, and can be used as an estimate.

To make sure that all errors have been corrected, Alice and Bob can verify whether their keys are identical. This verification process can be done by exchanging parity information \cite{Lutkenhaus1999, Fung2010}. Given $V$ parity sums announced from a key with a least one error, a very good approximation for the probability of an undetected error is 
\begin{equation}\label{parity}
p_{U\!|\!E}=\left(\frac{1}{2}\right)^V.
\end{equation}

As an alternative, we propose to exploit the minimum distance of LDPC codes as follows: After error correction Alice and Bob publicly announce $V$ randomly selected bit pairs. Since any non-identical keys have at least $d_{\text{min}}$ errors the probability of not finding any errors given that there exist some errors is given by
\begin{equation}\label{Undetected error}
p_{U\!|\!E}\leq\left(1-\frac{d_{\text{min}}}{N}\right)^V\leq(1-2\delta)^V
\end{equation}
This method is simpler and less computational demanding than exchanging parities, but more verification bits are needed to reach the same $p_{U\!|\!E}$.

If the actual error rate for a given block is larger than the estimate, Bob might end up with a wrong final key. Thus one should add a buffer $\Delta$ to the original estimate when running the protocol. The chosen value for $\Delta$ depends on the uncertainty in the error estimate and the consequences of coding into a wrong keyword. If the key only is used to encrypt information going from Alice to Bob, Bob having the wrong key only makes Alice message unreadable. On the other hand, Bob's key is not necessarily covered by security proofs if it differs from Alice's, so using it to encrypt data would be a breach of security.

We can now find expressions for the number of bits lost in error correction with error estimation by random sampling (EERS), and with verification. Assume that block $i$ has $N\delta_i$ errors. Let $\epsilon$ be an upper bound for the probability that the error correction step fails in a way such that Alice and Bob unwittingly end up with different codewords. This bound should be valid under any circumstances and arbitrary attacks by Eve. We assume that the error correction protocol used is based on LDPC codes, and for simplicity we assume that it performs at the Shannon limit for any error rate. 


\subsubsection*{Error estimation by random sampling (EERS):}
Random sampling of $S$ bits gives an estimated error rate $\delta_S$, which is approximately binomial distributed with mean $\delta_i$ and variance $\sigma^2_S=\frac{\delta_i(1-\delta_i)}{S}$. The loss in the error estimation and error correction is given by
\begin{equation} \label{Samplingloss}
L_{S}=S+(N-S)h(\delta_S+\Delta_S)
\end{equation}
with $\Delta_S$ being the buffer parameter. Assuming that sampling only makes a negligible change in the error rate of the $N-S$ remaining bits, the probability of an undetected error is bounded by 
\begin{align} 
p_U&=P(\delta_S+\Delta_S<\delta_i)\\\nonumber
&\leq\max_{\delta_i}P(\delta_S+\Delta_S<\delta_i).
\end{align} 
The maximization over all possible values for $\delta_i$ is necessary since we have no a priori information about the error rate. Using the normal distribution as an approximation for the binomial we get
\begin{align} \label{Sampling pue}
p_U&\leq\max_{\delta_i}\Phi\left(\frac{-\Delta_S}{\sigma_S}\right)\\\nonumber
&=\frac{1}{2}\left(1-\erf(\Delta_S\sqrt{2S})\right),
\end{align}
with $\Phi$ being the cumulative normal distribution function. A lower bound for $S$ such that \mbox{$P(\delta_S+\Delta_S<\delta_i)<\epsilon$} is then
\begin{equation}\label{Sampling}
S\geq\frac{1}{2}\left(\frac{\erf^{-1}(1-2\epsilon)}{\Delta_S}\right)^2 .
\end{equation}

\subsubsection*{Verification:}
Assume that Alice and Bob use the error rate of the previous block, $\delta_{i-1}$, plus a buffer parameter $\Delta_V$ as their estimate for the error rate. Assuming the worst case scenario, $p_{U|E}=\epsilon$, the loss is  given by
\begin{equation} \label{Verificationloss}
L_V=(p_E-\epsilon)N+(1-p_E+\epsilon)(V+Nh(\delta_{i-1}+\Delta_V))
\end{equation}
with $V$ being the number of bits used in verification step, and $p_E$ being the probability that Bobs raw key is transformed into the wrong codeword by the error correction protocol. 

Utilizing the minimum length between codewords the probability of an error not being detected is given by \eqref{Undetected error}. The probability of an undetected error is $p_U=p_{U\!|\!E}\,p_E$. Since we do not know Eve's action we have no certain knowledge about the block error rate $\delta_i$, and therefore we cannot bound $p_E$. Thus 
\begin{equation}p_U\leq\left(1-\frac{d_{\text{min}}}{N}\right)^V.\end{equation}
Note that this is independent of the actual error rate. Using \eqref{dmin} we find a lower bound for $V$ to ensure that $p_U\leq\epsilon$ to be
\begin{subequations}\label{Verification}
\begin{equation}
V\geq\frac{\log(\epsilon)}{\log(1-\frac{d_{\text{min}}}{N})}=\frac{\log(\epsilon)}{\log(1-2(\delta_{V}+\Delta_V))}.
\end{equation}
As noted we can also do the verification by parity exchange. The number of bits used in verification is then, using \eqref{parity},
\begin{equation}
 V\geq\frac{\log(\epsilon)}{\log(\frac{1}{2})}.
\end{equation}
 \end{subequations}

For a system where every bit has the same a priori probability of being an error, $\delta_i$ and $\delta_{i-1}$ are both normally distributed with mean $\delta$ and variance $\sigma^2$. In that case we have
\begin{equation}
p_E=P(\delta_i\geq\delta_{V}+\Delta_V)=\Phi\left(\frac{-\Delta_V}{\sqrt{2}\sigma}\right),
\end{equation}
which we can use in \eqref{Verificationloss} to find the total loss.

\section*{Numerical results}
For performance analysis, we first consider a system running with mean error rate $\delta$ and variance between block error rates $\sigma^2=\frac{\delta(1-\delta)}{N}$, i.e. all variance is due to the inherit randomness of the bit values. The loss in error correction is then dependent on three parameters, the error rate $\delta$, the security parameter $\epsilon$ and the block size $N$. 

We can minimize the loss from the error correction, \eqref{Samplingloss} and \eqref {Verificationloss}, for different $\delta$, $\epsilon$ and $N$, with respect to the buffer parameters $\Delta_j,\ j=S,V$. Note that when running the error correction protocol, the value of the buffer parameter is chosen according to the estimates $\delta_S$ and $\delta_{V}$, not the error rate $\delta$. Since these estimates are not exact, one will generally choose a suboptimal value for $\Delta_j$, resulting in slightly larger losses than the one showed in the following results. Also note that the possibility of choosing a suboptimal value for $\Delta_j$ is accounted for in security analyses in the previous section.

We define the excessive loss ratio, $L^E_j$, to be
\begin{equation} \label{Excessiveloss}
L^{E}_j=\frac{L_j}{N}-h(\delta)\qquad j=S,V 
\end{equation}
\begin{figure}
\includegraphics [width=0.48\textwidth]{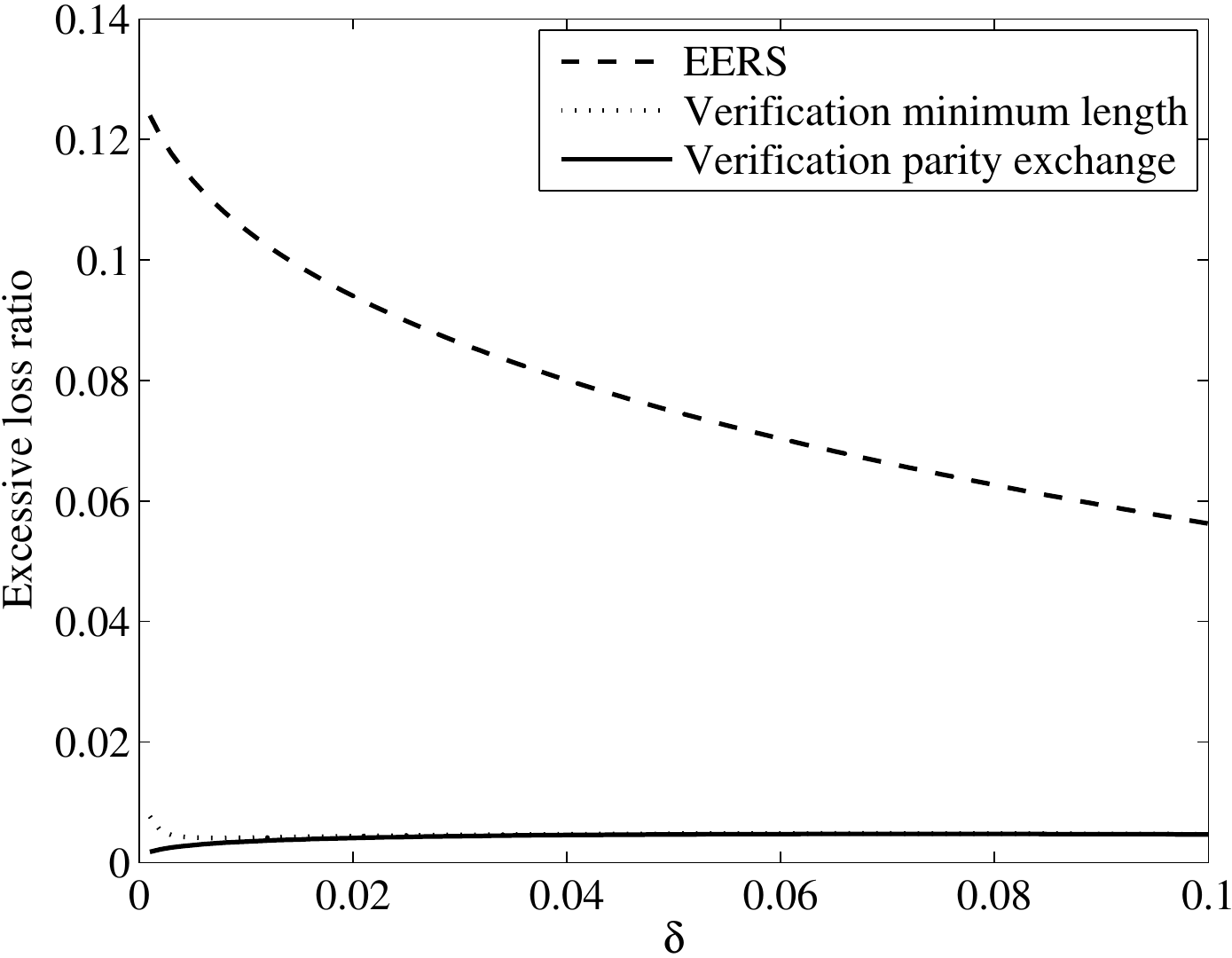}
\caption{\label{delta} Loss ratio for different error rates.  $N=10^6$, $\epsilon=10^{-6}$.}
\end{figure}
\begin{figure}
\includegraphics [width=0.48\textwidth]{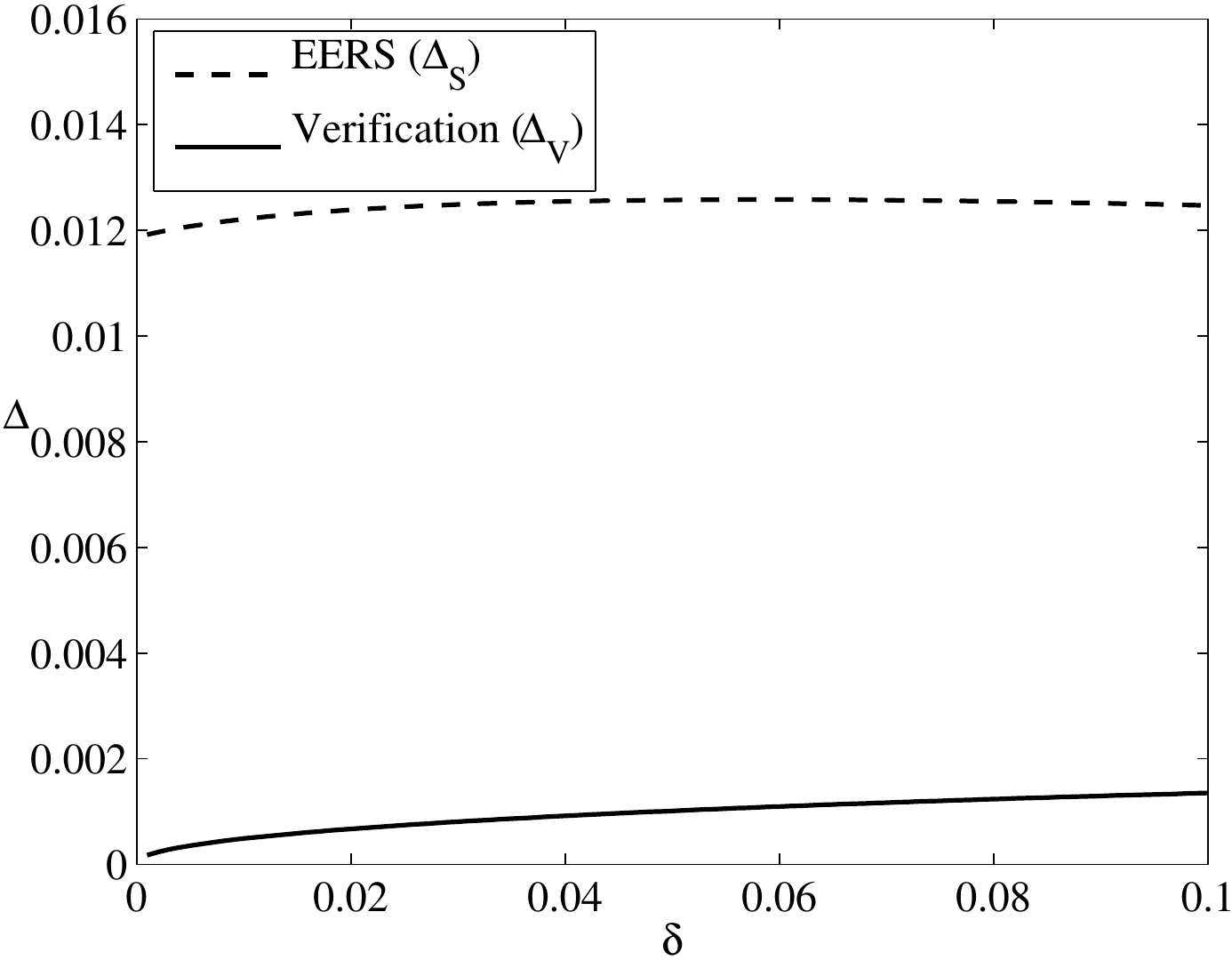}
\caption{\label{Delta} Optimal value of the buffer parameter $\Delta$ for different error rates.  $N=10^6$, $\epsilon=10^{-6}$.}
\end{figure}
Figure \ref{delta} shows that the excessive loss ratio is lower for verification than for EERS for all error rates $\delta$. We also see that the difference between the two methods of verification is small compared to the difference between error correction and verification, especially for large $\delta$. Since the difference is close to negligible we consider verification by utilizing minimum length between codewords in the rest of the discussion. All results also apply to verification by parity exchange unless noted otherwise.

There are two main terms contributing to the difference, both related to the security parameter $\epsilon$. As mentioned, the probability of undetected errors, bounded by $\epsilon$, might be of critical importance to the security of the protocol. If we use error estimation we must have a high buffer parameter $\Delta_S$ to avoid such errors. However, if we use verification, we have an efficient method to find errors after the error correction. The main purpose of $\Delta_V$ is then not to avoid all errors, but only to keep the error probability $p_E$ low to avoid many blocks being thrown away. We can then choose a buffer parameter $\Delta_V<\Delta_S$ even though our estimate $\delta_{i-1}$ is less reliable than $\delta_S$. Optimal values for $\Delta_V$ and $\Delta_S$ are shown in Figure \ref{Delta}. 

\begin{figure}
\includegraphics [width=0.48\textwidth]{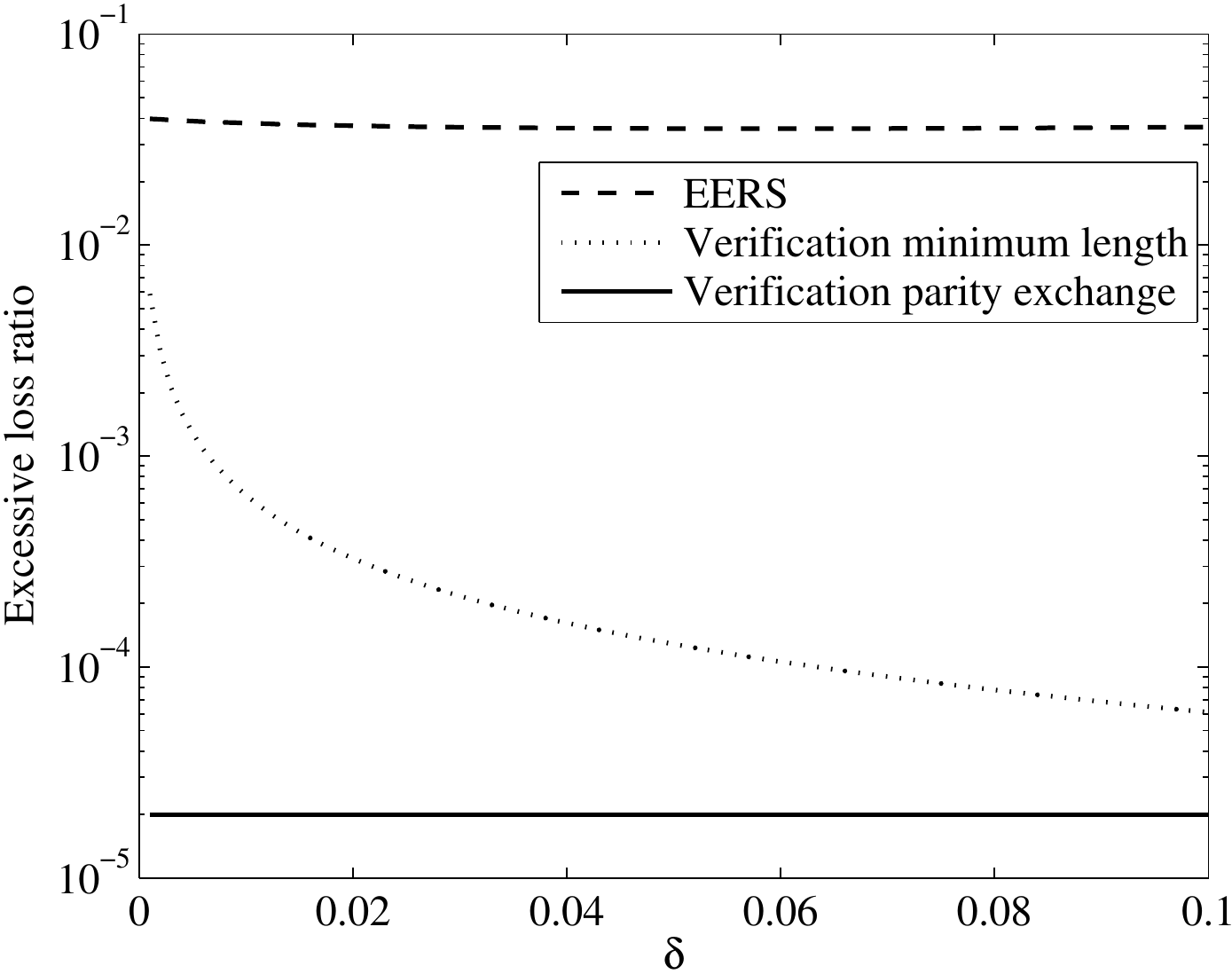}
\caption{\label{samplingbits} Excessive loss ratio from the sampling procedure.  $N=10^6$, $\epsilon=10^{-6}$.}
\end{figure}
The other reason that verification has a smaller excessive loss than EERS is that to keep $\Delta_S$ from growing too large we must use a large sample size $S$. This sample is much larger than the number of bits $V$ used for verification. Actually, as seen in Figure \ref{samplingbits}, $V$ does not give a significant contribution to the excessive loss unless we are using the minimal length approach on a raw key with very small $\delta$. This again shows that the method one chooses for verification, exchange of parities or utilizing the minimum length between codewords, is not important when it comes to excessive loss unless $\delta$ is very small. 

%

\begin{figure}
\includegraphics [width=0.48\textwidth]{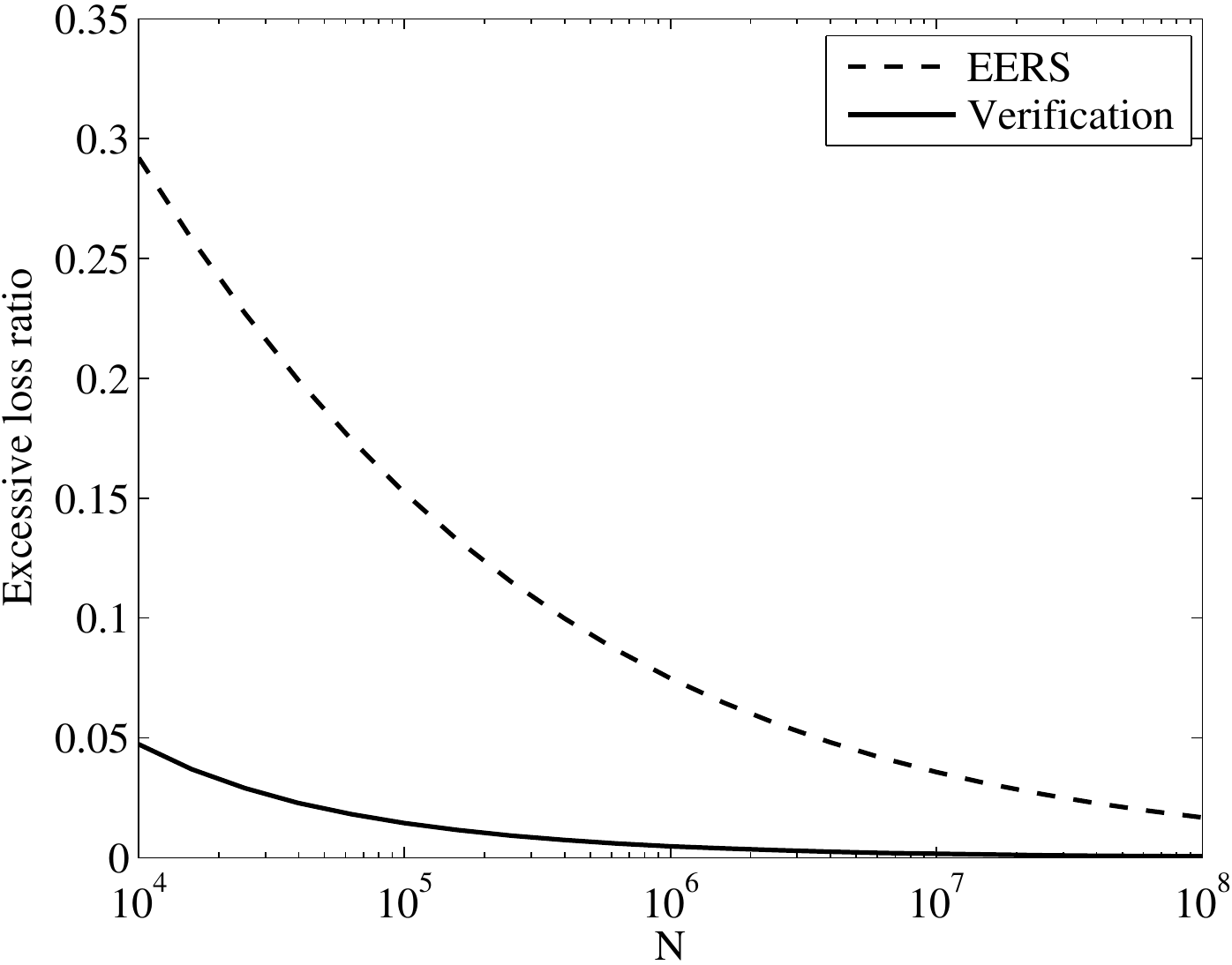}
\caption{\label{blocksize} Loss ratio for different block size.  $\delta=0.05$, $\epsilon=10^{-6}$.}
\end{figure}
As shown in Figure \ref{blocksize} the block size $N$ is crucial to the excessive loss ratio. For EERS the high loss ratio for small $N$ is mainly due to a large part of the block being used in the sampling process. Using verification this loss is avoided. Here the increased loss ratio for small $N$ is due to the larger variance between block error rates when $N$ is small. 

\begin{figure}
\includegraphics [width=0.48\textwidth]{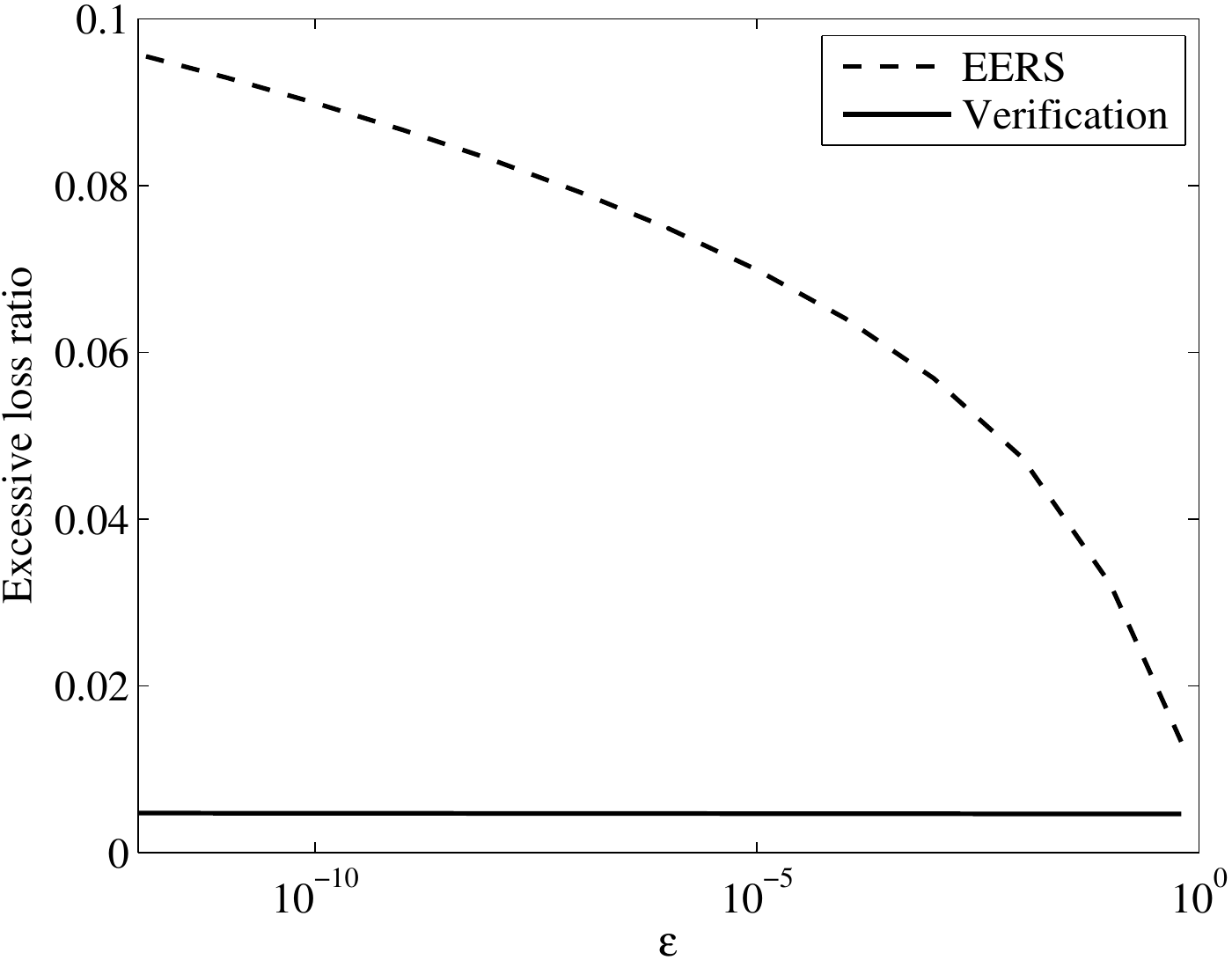}
\caption{\label{epsilon} Loss ratio for different security parameters $\epsilon$.  $\delta=0.05$, $N=10^6$.}
\end{figure}
In verification, better security, i.e. decreasing the security parameter $\epsilon$, demands more bits $V$ used to check for error after the error correction. However since $V\sim\log{\epsilon}$ \eqref{Verification}, and additionally $V\ll Nh(\delta_V+\Delta_V)$, decreasing $\epsilon$ only gives a minimal increase in the loss ratio \eqref{Verificationloss}. Thus, as shown in Figure \ref{epsilon}, we can increase the security tremendously while sacrificing few extra bits if we use verification. In the scheme of EERS, as $\epsilon \rightarrow 0$, $S$ increases towards infinity quite fast because of the inverse error function in \eqref{Sampling}. Since sampling is a significant part of the loss ratio for all but very large $N$, high security comes with a high excessive loss ratio in this scheme.

\section*{Variable error rates}
\begin{figure}
\includegraphics [width=0.48\textwidth]{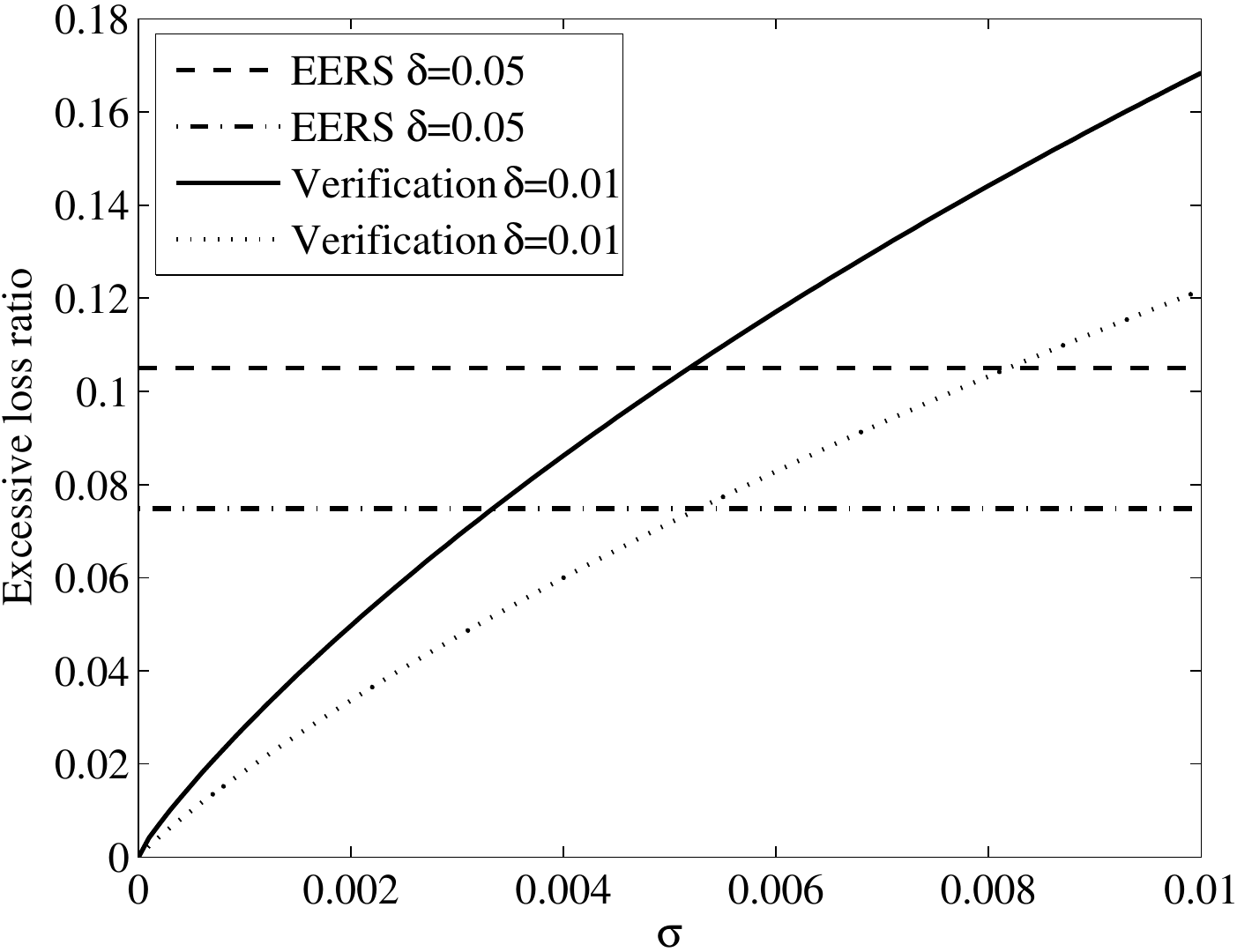}
\caption{\label{sigma} Excessive loss ratio for different block error rates and block error rate variance. The block error rates are assumed to be independent and normally distributed. $N=10^6$, $\epsilon=10^{-6}$.}
\end{figure}

In real setups external factors like temperature fluctuations and calibration routines may cause greater variation in the block error rate. Then, using the error rate of the last block as our estimate for the error rate of the current block, is less reliable. To avoid throwing away more blocks due to the less accurate estimates, the buffer parameter, $\Delta_V$, must be increased. This will lead to increased loss in the protocol. Using an EERS scheme, the loss is independent of the block error rate variance. Thus, as shown in Figure \ref{sigma}, verification is preferable when the block error rate variance is small, while EERS should be considered when the  variance is high.

\begin{figure}
\includegraphics [width=0.48\textwidth]{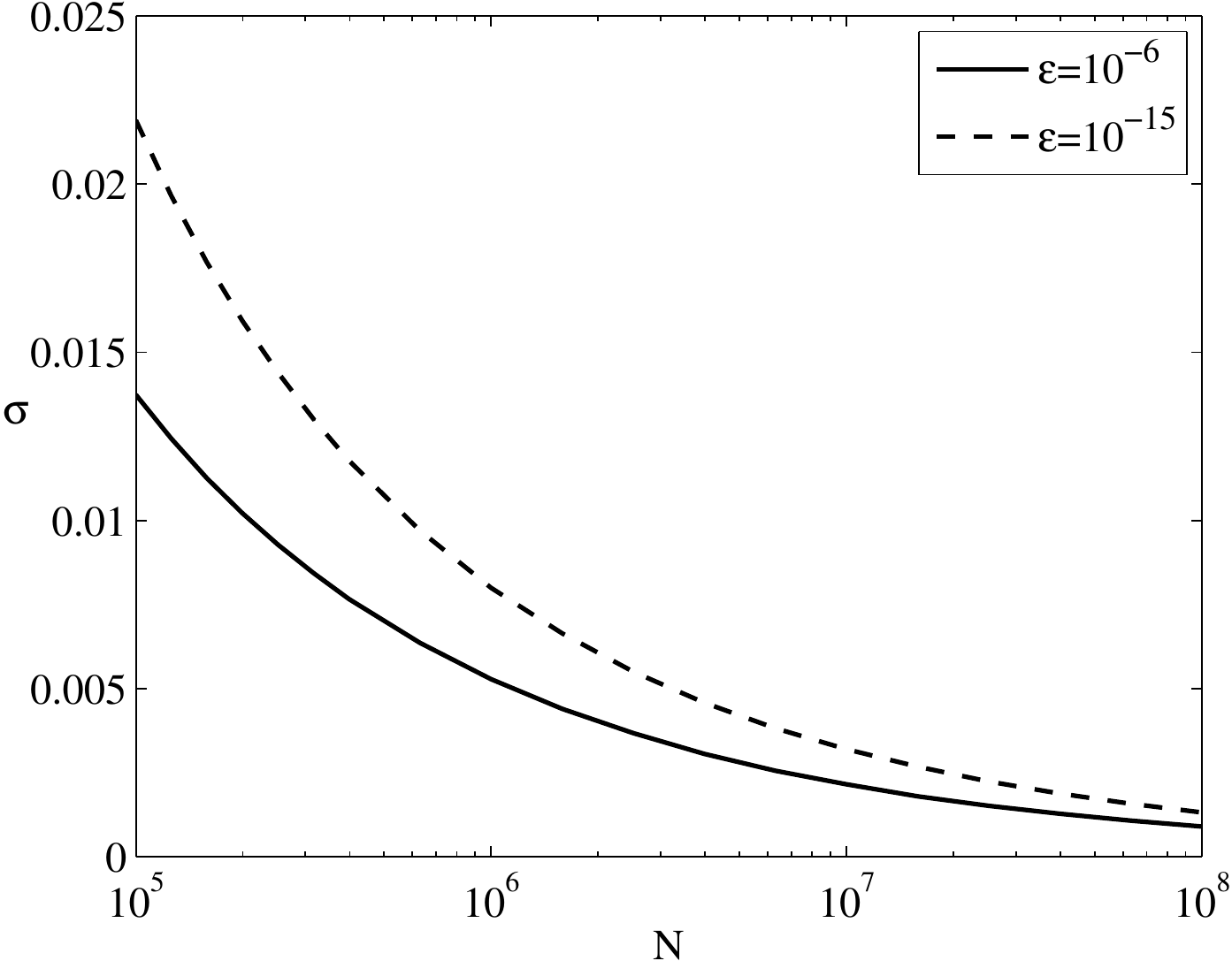}
\caption{\label{sigmaNepsilon} The curves show for which block error rate variance and block size EERS and verification has the same excessive loss. Verification is the best method for parameters in the area to the left of the curves, while EERS is best for parameters to the right of the curves. $\delta=0.05$.}
\end{figure}

Figure \ref{sigma} also indicates that the variance for which sampling and verification has equal excessive loss only depends slightly on $\delta$. Thus the important variables are $N$ and $\epsilon$. As shown in Figure \ref{sigmaNepsilon} large variance favors EERS while small block size and high security demands favor verification.

\begin{figure} 
 \includegraphics [width=0.48\textwidth]{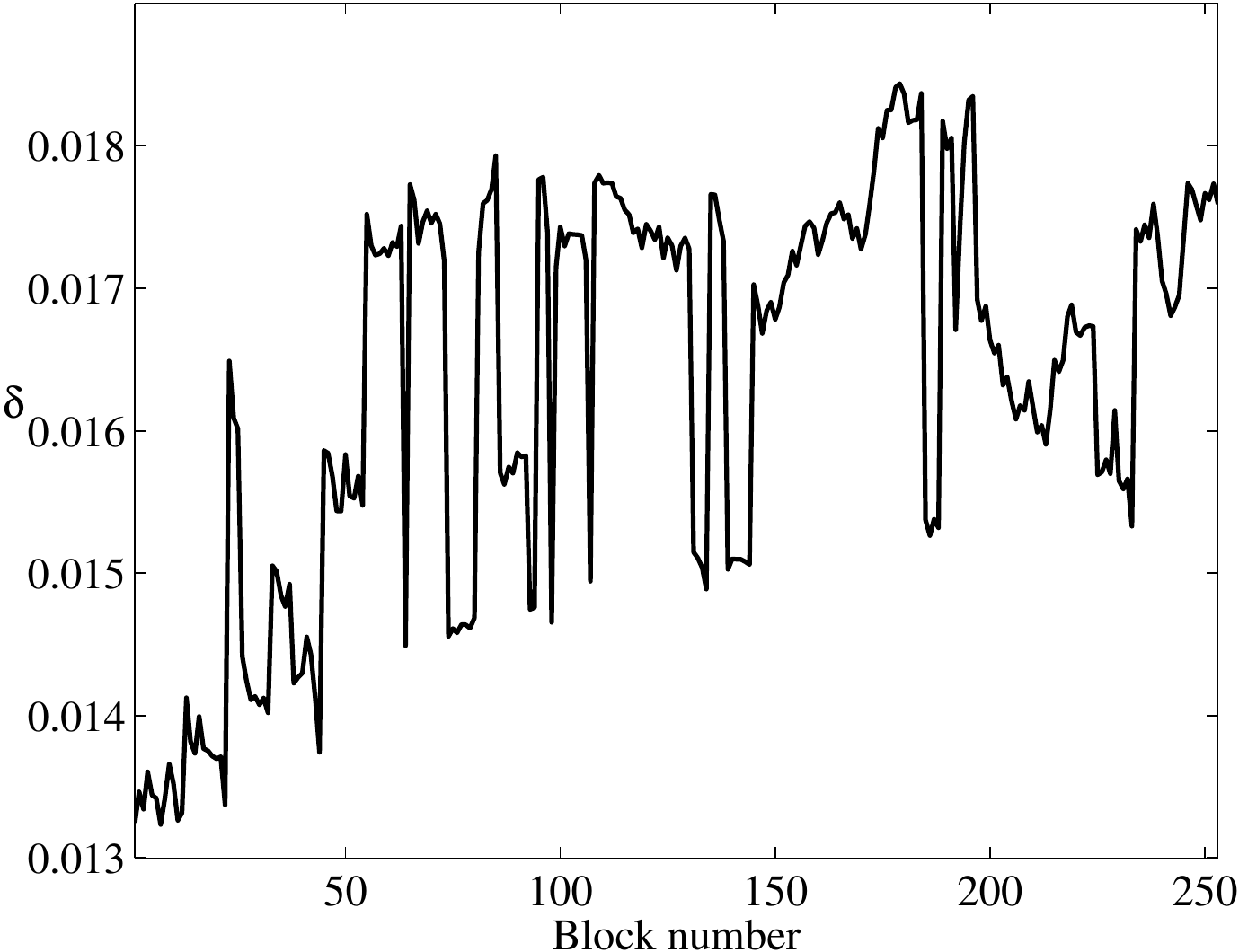}
 \caption {\label{errorrate} Block error rate for a 24-hour run of the IdQuantique system Clavis$^2$}
\end{figure}

In real setups the block error rate is not necessarily normally distributed. For example, Figure \ref{errorrate} shows how the block error rate evolved for a 24-hour run of the IdQuantique system Clavis$^2$. In this case it is difficult to model the block error rate and thus to find an optimal $\Delta_j$. However, as can be seen from the figure, in this run $\Delta_V=0.004$ would be enough to avoid any errors. Minimizing \eqref{Samplingloss} with respect to $\Delta_S$ for the relevant $N=2.6\cdot10^6$, $\delta\approx 0.016$, and $\epsilon=10^{-6}$  we find the optimal buffer parameter for EERS to be $\Delta_S\approx 0.009$. Thus it seems that verification would give the smallest excessive loss for this setup. Calculating the actual values we find $L^E_V=0.023$ and $L^E_S=0.074$.
However, this is only true as long as it continues its current behavior. If the variance in the block error rate changes so does the optimal buffer parameter and maybe also the optimal method. 

In fact one of the assumption used in calculating these results, that we always manage to choose the buffer parameter $\Delta_j$ close to its optimal value, might not be justified for the verification scheme if the block error rate start to fluctuate in an unexpected way. Then there is a risk of having loss much larger then expected. The EERS scheme is not prone to this problem since $\Delta_S$ might be estimated pretty accurately from $\delta_S$. Thus EERS is recommended for systems with unknown behavior. In this respect the IdQuantique system the seems quite stable. Considering groups of 50 consecutive blocks, the error rate between blocks varies a lot within each group. However the distribution of the difference between each block is quite similar for all the groups. Especially is the maximal difference between two consecutive blocks, which is the important quantity in finding a good value for $\Delta_V$, very similar in all the groups. Thus it seems that verification scheme with $\Delta_V=0.004$ would 
work fine also for the next 250 blocks.

For the 24-hour run of the IDQuntique system 30.9 percent of the raw key was lost in error correction, mostly due to whole blocks beeing discarded. This gives an excessive loss ratio of 0.189. It clear that a better error correction scheme would be beneficial to the systems performance.

\subsubsection*{Combination of the methods}
We have seen that using EERS many bits must be sacrificed in random sampling to achieve high security. On the other hand, when the variance in the block error rate is high, doing verification and using the previous block as an estimate for the error rate also has large excessive loss since the estimate is not very accurate. Thus, if the block error rate variance is high and we want high security combining the two methods make sense.

The loss related to error correction using both EERS and verification is, again assuming $p_{U|E}=\epsilon$,
\begin{align}
L_C=&(p_E-\epsilon)N+(1-p_E+\epsilon)\\\nonumber
&\cdot(S+V+(N-S)h(\delta_C+\Delta_C)).
\end{align}
Just like for EERS the loss is independent of the variance and the method is robust against wild fluctuations in the block error rate. 

Using the results from the EERS as our estimate $\delta_C$, the probability $p_E$ of an error after the error correction step is the same as the probability given in \eqref{Sampling pue} with $\Delta_C$ for $\Delta_S$. Using verification by parity exchange the probability of an undetected error is then
\begin{equation}
 p_U=p_{U\!|\!E}\,p_E=\left(\frac{1}{2}\right)^{V+1}\left(1-\erf(\Delta_C\sqrt{2S})\right).
\end{equation}

For a given security parameter $\epsilon$ the number of bits used in error estimation is then related to the bits used in verification by
\begin{equation}
 V=\log(1-\erf(\Delta\sqrt{2S}))-\log\epsilon-1
\end{equation}
We define the excessive loss $L^E_C$ as in \eqref{Excessiveloss} with $j=C$. This can now be minimized with respect to the buffer parameter $\Delta_C$ and the sampling size $S$.

For $\epsilon=10^{-6}$ and $N=10^6$ the results are shown in Table \ref{Combination results}. 
\begin{table}[t]
\resizebox{0.48\textwidth}{!}{
    \begin{tabular}{|l|l|l|l|l|} 
    \hline &&&&\\[-2ex]
	Error rate     			& Method      & $\Delta$ & $\tfrac{S+V}{N}$ & $L^E$ \\ 
	&&&&\\[-2.5ex]\hline
        \multirow{2}{*}{$\delta=0.05$}	& EERS        & 0.0126	& 0.036 & 0.075  \\ 
        ~           			& Combination & 0.0081	& 0.023 & 0.053  \\ \hline
        \multirow{2}{*}{$\delta=0.01$} & EERS        & 0.0122	& 0.038 & 0.105  \\ 
        ~           			& Combination & 0.0077	& 0.025 & 0.076  \\
     \hline  
    \end{tabular}}
\caption{\label{Combination results}Results for EERS and a combination of EERS and verification.}
\end{table}
We clearly see that using a combination of the methods leads to an improvement in performance compared to EERS alone. We expect this improvement to be even more profound if we demand higher security (decreases $\epsilon$), or for small block sizes, as these are scenarios where verification significantly outperforms EERS.

To compare the combination method with verification we can compare the results from Table \ref{Combination results} with Figure \ref{sigma}. As the performance of the combination method is independent of variance we infer that it outperforms verification when the variance is larger than 0.004 while verification is better for $\sigma<0.003$. 

Going back to the block error rate from the IdQuantique system we find $L^E_C=0.054$ for combination of the methods. Thus the variance between block error rates is so small that it seems verification only is the best approach for this system. 

\section*{Conclusion}
Due to the uncertainty about the true value of the block error rate some bits need to be sacrificed to decrease the probability that Alice and Bob have undetected errors in their keys. This can be done by EERS before the error correction protocol, or by verification after the protocol. We find that verification generally outperforms EERS, however if the variance in the block error rate is large EERS is the best choice. To minimize the loss in error correction it is therefore important to have a QKD system with a stable error rate. 

We propose a combination of the two methods that generally outperforms EERS. This combination method, and EERS, are both robust against changes in the behavior of the error rate. If one only does verification, large losses might occur if the block error rate changes unexpectingly. Thus the combination method should be used when the variance of the block error rate is high or when the change in the error rate between blocks is unknown or susceptible to unpredictable fluctuations.

We also show that utilizing the minimum distance of LDPC codes provides a fast and efficient way to do verification.

\bibliography{bibtex_library}
\bibliographystyle{ieeetr}

\end{document}